\documentclass{article}
\usepackage{ijcai18}

\usepackage{times}
\usepackage{xcolor}
\usepackage{soul}
\usepackage[utf8]{inputenc}
\usepackage[small]{caption}

\usepackage{graphicx}
\usepackage{subcaption}

\graphicspath{{images/}}
\usepackage{todonotes}
\usepackage[nointegrals]{wasysym}
\usepackage{amsmath}
\usepackage{float}
\usepackage{booktabs}

\usepackage{comment}

\title{An agent-based model of an endangered population of the Arctic fox from Mednyi Island}

\author{
Angelina Brilliantova$^1$, 
Anton Pletenev$^1$, 
Liliya Doronina$^2$,
Hadi Hosseini$^3$
\\ 
$^1$ Moscow State University, Russia \\
$^2$ University of Münster, Germany  \\
$^3$ Rochester Institute of Technology, USA\\
cheli231@gmail.com,
aapletenev@yandex.ru,
doronina@uni-muenster.de,
hhvcs@rit.edu
}

\begin{document}

\maketitle

\begin{abstract}
Artificial Intelligence techniques such as agent-based modeling and probabilistic reasoning have shown promise in modeling complex biological systems and testing ecological hypotheses through simulation. We develop an agent-based model of Arctic foxes from Medniy Island while utilizing Probabilistic Graphical Models to capture the conditional dependencies between the random variables. Such models provide valuable insights in analyzing factors behind catastrophic degradation of this population and in revealing evolutionary mechanisms of its persistence in high-density environment. 
Using empirical data from studies in Medniy Island, we create a realistic model of Arctic foxes as agents, and study their survival and population dynamics under a variety of conditions. 

\end{abstract}

\section{Introduction}

Studying demographic mechanisms of a population for conservation risk assessment is often a challenging issue since ecological systems are inherently complex, nonlinear, and include multiple interactions of various components \cite{witmer2005wildlife}. The complex nature of ecological, biological, and behavioural aspects of species calls for novel interdisciplinary approaches for understanding the subtle reasons for survival and extinctions of species. In recent years, Artificial Intelligence (AI) techniques such as Agent-Based Modeling (ABM) have shown promise in modeling complex systems under uncertainty, analyzing their characteristics, and allowing the testing of more complex biological and ecological hypotheses through simulation \cite{grimm2013individual,wellman2016putting,mclane2011role,yang2014adaptive}. Agent-based modeling have become widely used for addressing eco-evolutionary \cite{mosser2015landscape} and epidemiological \cite{eisinger2008spatial,wang2015exploring} issues, in particular, for studying population dynamics and improving conservation management. \cite{robbins2004simulation,stenglein2015individual,carter2015modeling,watkins2015spatially}.
The investigation of population dynamics in endangered Arctic fox subspecies (\textit{Vulpes lagopus semenovi}, listed in Russian Red Data Book, \cite{goltsman1996disease}) from Mednyi Island (Commander Islands, North Pacific) is one of the intriguing examples that requires application of modern AI approaches. Until the end of the 20th century, the population density remained extremely high as compared to other fox populations: up to a thousand animals on an island of 187 square kilometers. An epizootic of ear mange occurred in 1970--1980 among juveniles that wiped out the majority of cubs and led to a drastic decline in a population size. During the next few decades the population had partially recovered, however, it stabilized on a much lower level, hovering around 100 adults (10-15\% of the previous size) \cite{goltsman2005effects,goltsmanIsland}.

To study the effects of several factors on the population dynamics we develop an agent-based model of the population that incorporates uncertainties through probabilistic variables. In this model, Arctic foxes are represented as agents with particular attributes, interests, and behaviors that can interact with other agents and the environment.
This probabilistic model sheds light into factors behind catastrophic degradation of the Arctic fox population and reveals evolutionary mechanisms allowed for the population persistence in high-density environments.
Given the data collected between 1994--2012, we investigate various factors in survival of Arctic fox population by empirically studying various scenarios and the interaction between parameters that affect the population. 
In future, this model can provide ecologists with a powerful tool for population level prediction, effective protection, and management of the ecosystems.
 

%


\section{Background}

The demography of Mednyi Arctic foxes has been extensively investigated throughout recent decades. From 1994 until 2012 approximately 80\% of arctic foxes living on the southern part of Mednyi Island were marked with plastic ear-tags and were individually recognized during their lifetime. In result, all basic life-history parameters of the island population were determined \cite{goltsman1996disease,goltsmanIsland,kruchenkova2009alloparenting,doroninaprep}. Based on this long-term individual-based field study the attempts to model the population dynamics were made. The results of matrix modeling \cite{matrixMednyi2017} showed that the population growth rate most notably depends on juvenile survival. Data from tagged animals was used to develop first agent-based model of Mednyi Arctic fox population \cite{ABM2018}. The model presented in this study differs from \cite{ABM2018} in explicitly incorporating probabilistic conditional framework which allows for future modification and expansion of the model.

\section{The Model}

\begin{figure}
    \centering
    \caption{Probabilistic graphical model of the Arctic foxes. Widely supported primary dependencies are depicted in black.}
\begin{subfigure}[]{0.42\textwidth}
\caption{Complete probabilistic model of the Arctic foxes}
\centering
\label{fig:pgm}
\includegraphics[width=\textwidth]{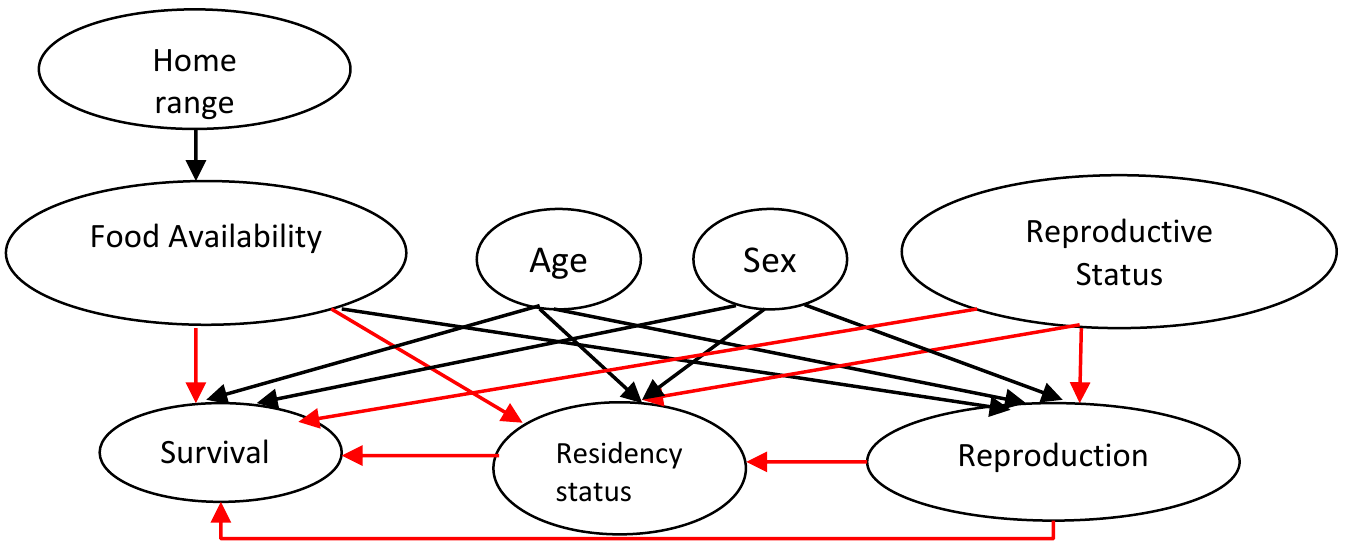}
\end{subfigure}
\begin{subfigure}[]{0.42\textwidth}
\caption{Probabilistic model of the Arctic foxes}
\centering
\label{fig:pgm_simple}
\includegraphics[width=\textwidth]{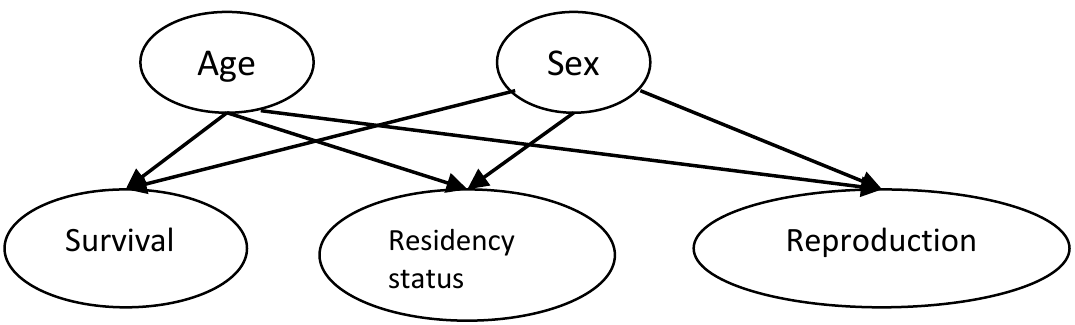}
\end{subfigure}
\end{figure}

This section introduces the formalism required to model the ecosystem. Our agent-based model consists of three primary components: agents, the environment, and processes.

\textbf{Agents:}
In our model, each arctic fox is modeled as an agent $i$ with type $\theta_{i} = (s_{i}, a_{i}, r_{i}, rp_{i}, h_{i})$ where $s_{i}\in \{f,m\}$ indicates the gender, $a_{i}$ denotes the age, $r_{i}\in \{0,1\}$ is a boolean variable denoting the residency status with $0$ representing floaters and $1$ resident agents, $rp_{i} \in \{$\text {breeder}, \text{helper}, \text{member of non reproductive group}, \text{peripheral animal}, \text{single animal}$\}$ is a categorical variable denoting the reproductive status of agent $i$, and $h_{i}$ denotes a home range occupied by the agent $i$.

The development stages of an agent is modeled by a linear function over time. All animals who survive in each period grow for $1$ year in each time period $t$, thus, $a_{i}^{t+1} = a_{i}^{t} + 1, \forall a_{i}^{t}$.
Following the literature on the lifespan of arctic foxes, we set up a limitation for a $max (a_{i}^{t}) = 12$ which exceeds the maximum observed age of Arctic foxes, 10 years, \cite{hersteinsson1992demography,eide2012reproductive}. Therefore, agents with $a_{i}^{t} > 12$ are eliminated at each time $t$.
We further classify the developmental stages as age classes $A_{i}$:
\begin{gather}
\small
  A_{i}=\left\{
  \begin{array}{@{}ll@{}}
    cub, & \text{if}\ a_{i}=0 \\
    yearling, & \text{if}\ a_{i}=1 \\
    adult, & \text{if}\ a_{i} \in [2, 12]
  \end{array}\right.
\end{gather} 

\textbf{Environment:}
The landmark of the island can be considered as a collection of territories or home ranges. Let $H^{t}$ denote the set of home ranges at time $t$. Each home range $H_{j} \in H^{t}$ has a set of attributes, that is, 
$H_{j} = (x_{j}, n_{j}^{t}, n_{j}^{m,t}, n_{j}^{f,t}, f_{j})$, where $x_{j}$ is the coordinate of home range $j$, and $n_{j}^{t}, n_{j}^{m,t}, n_{j}^{f,t}$ are inhabitancy parameters indicating the total number of agents, the number of agents with $s_{i} = m $, number of agents with $s_{i} = f $ inhabiting $H_{j}$ at time $t$ respectively; in all cases the number of agents does not account for cubs. Thus, for all $j, t$ we have $n_{j}^{t} = n_{j}^{m,t} + n_{j}^{f,t}$. Each $H_{j} \in H^{t}$ has a corresponding food availability level $f_{j}\in \{poor,medium,rich\} $, based on proximity and size of bird colonies near location $x_{j}$. For simplicity, in this paper the configuration of home ranges ($H^{t}$) along the island is assumed identical for all periods, that is, $H_{j}^{t} = H_{j}^{t'}, \forall t,t'$. 
The total population at time $t$ is denoted by $n^{t}$, where $n^{t} = \sum_{H_{j}\in H^{t}} n_{j}^{t}$, thus, $n^{0}$ is the initial population at $t = 0$. 


\subsubsection{A Probabilistic Life Cycle}
The life cycle of animals is heavily influenced by uncertainties derived from a variety of environmental factors. For example, in Mednyi Arctic fox population, survival and dispersal are strongly affected by the sex of animals, and the sex ratio in litters is influenced by food availability in a home range: in rich habitats females produce more females, while in poor conditions the sex ratio of cubs is biased towards males \cite{goltsmanIsland,goltsman2005effects}. 

To model the inherent conditional dependencies between the random variables, we utilize Probabilistic Graphical Models (PGM) for the Mednyi Arctic fox population (Fig.\ref{fig:pgm}). PGMs are probabilistic models illustrating structured conditional dependencies between random variables in complex domains. 
In Fig.\ref{fig:pgm} several likely dependencies are depicted with red links as the current literature on arctic foxes has yet to explore the causal effects. Therefore, we focus attention on primary dependencies that are widely supported in the literature \cite{goltsmanIsland,goltsman2005effects}, particularly on the affects of Age ($A$) and Sex ($s$) on survival, residency status, and reproduction (Fig.\ref{fig:pgm_simple}). The $f$ and $rp$ parameters are excluded in our current model for simplicity.

\textbf{Survival:}
Let $\phi_{i}$ denote the output of survival process for agent $i$ where $1$ denotes survival and $0$ denotes death of $i$. Survival process follows a Bernoulli distribution with $p(\phi_{i}=1|A_{i}, s_{i})$ as a probability of success. For each agent $i$, the age- and sex-specific survival is computed using Bayes rule. We assume that $A_{i}$ and $s_{i}$ are independent variables, so we can write:
\small
\begin{align*}
p(\phi_{i}|A_{i}, s_{i}) =  \frac{p(A_{i}, s_{i}|\phi_{i}) \times p(\phi_{i})}{p(A_{i}, s_{i})} 
=  \frac{p(A_{i}|\phi_{i}) \times  p(s_{i}|\phi_{i})\times  p(\phi_{i})}{p(A_{i})\times  p(s_{i})}
\end{align*}
\normalsize
where marginal probabilities of age class $p(A_{i})$ and sex $p(s_{i})$ are estimated using observed data as the ratio of agents in each age class and sex in a population. Conditional probabilities of age class $p(A_{i}|\phi_{i})$ and sex $p(s_{i}|\phi_{i})$ given survival output are estimated as proportions of agents with corresponding age and sex attributes separately in a sample of survivors and dead agents. 
Although, $p(\phi_{i}|A_{i}, s_{i})$ can be estimated directly from the empirical data, we used Bayes rule (1) to allow for future sensitivity analysis of age- sex-specific survival to the components of the right hand side of the equation, (2) to provide general framework for the future development of a complete PGM model (i.e. adding missing factors from Fig.\ref{fig:pgm} for which empirical data is not always available). 

\begin{figure}
\caption{Dispersal algorithm}
\centering
\label{fig:dispersal}
\includegraphics[width=.33\textwidth]{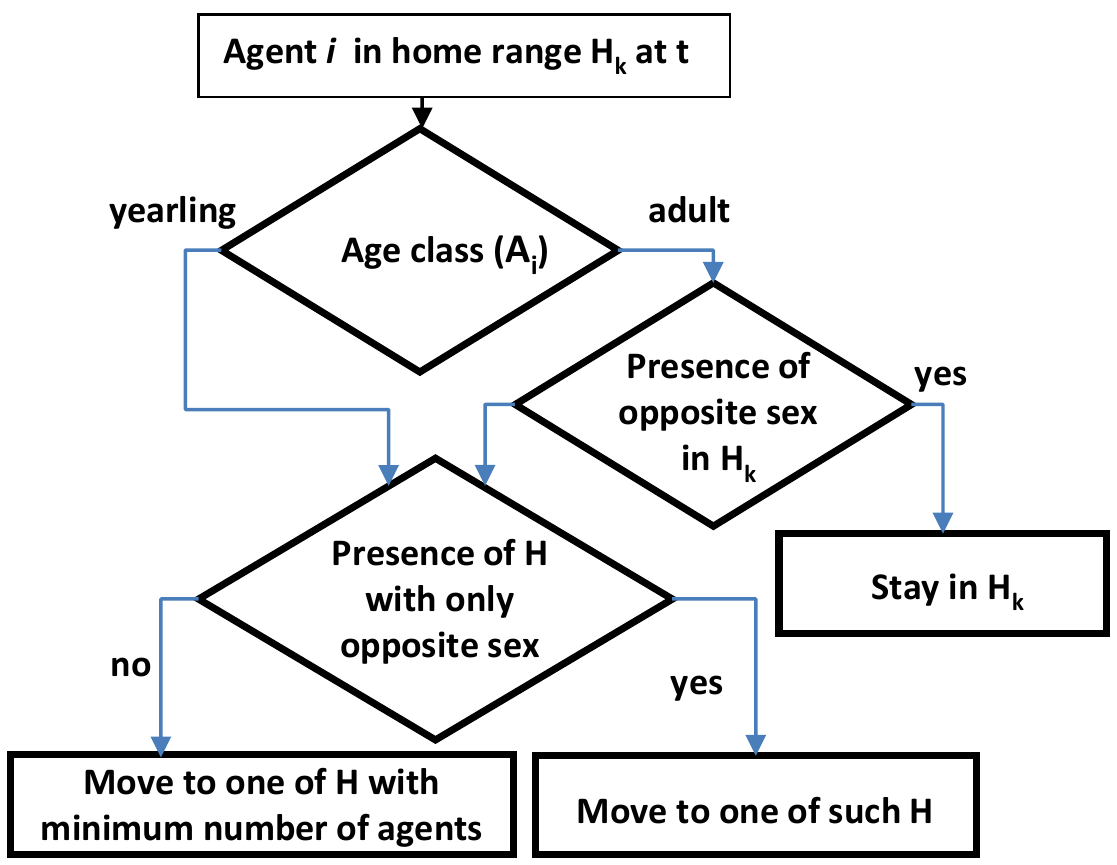}
\end{figure}

\textbf{Residency status and dispersal process:}
During dispersal, yearlings and lone adults change their home range trying to find a mate. For each agent the cycle checks its residency status. Let agent ${i}$ occupies ${H_{k}^{t}}$ at $t$. If $A_{i}^{t} = adult$ and $n_{k}^{sex	\neq  s_{i}, t} > 0$, i.e. there are more than 0 animals of the opposite sex in $H_{k}^{t}$, then agent $i$ is classified as a resident with $r_{i}=1$. Otherwise, the agent is classified as a floater with $r_{i}=0$ and will be forced to change its home range according to the following rule: If the set of $H^{t}$ with $n^{sex	\neq  s_{i}, t} > 0$ and  $n^{sex	= s_{i},t} =0$ is non-empty, then the agent moves to a randomly selected $H_{z} \in H^{t}$. Otherwise, the agent randomly moves to $H_{z}$ with $min(n_{j}^{t})$, $\forall {j}$. After dispersal, parameters $n^{t},n^{m,t},n^{f,t}$ of $H_{k}^{t}$ and $H_{z}^{t}$ are updated (see Fig.\ref{fig:dispersal}).

\textbf{Reproduction:}
Agents with $s_{i} = f$ and $A_{i}=adult$ or $yearling$ inhabited $H_{k}^{t}$ with $n_{k}^{m,t} > 0$ reproduce with probability $p^{r,a}$ for $A_{i} = adult$ and $p^{r,y}$ for $A_{i} = yearling$. We assume these probabilities are drawn i.i.d from a fixed distribution. The litter size is drawn from a normal distribution with mean equals $b$ and standard deviation equals $\sigma_{b}$. A newborn agent $i$ has $a_{i}=0$, $A_{i}=cub$, $h_{i} = H_{k}$, $r_{i} = 0$, and $s_{i}$ is drawn from a Bernoulli distribution with probability $p_{sex}$.

\section{Process overview and parameters}
\begin{figure}
\caption{The transitions between different age classes. Diamonds denote conditional and rectangles indicate stochastic processes.}
\centering
\label{fig:lifecycle}
\includegraphics[width=.44\textwidth]{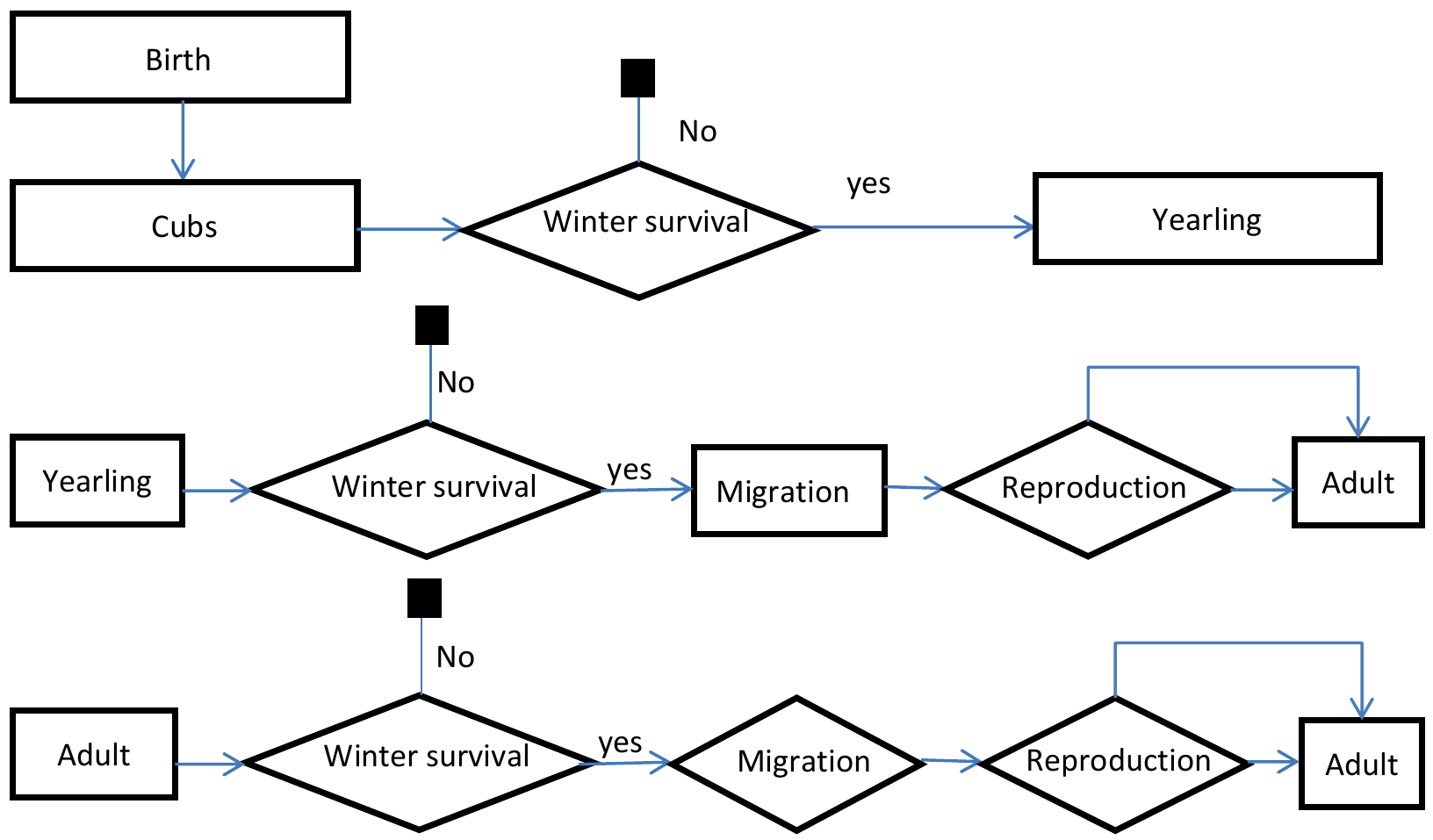}
\end{figure}

The model proceeds in annual time steps, starting from winter. Within each year or time step, three modules or phases are processed in the following order: winter survival, dispersal, and reproduction. Within each module, individuals and territories are processed in a random order. The individual life cycle is depicted in (Fig.\ref{fig:lifecycle}).
The model parameters were defined according to the field studies conducted on Mednyi Island between 1994--2012 (Table \ref{tab:default}). 
The survival parameters are estimated based on the dataset of life stages of animals tagged with individual ear tags when they were cubs. The total number of tagged cubs is 517.

\begin{table}[]
\footnotesize
\centering
\caption{The parameters and default values of the model}
\label{tab:default}
\begin{tabular}{@{}lllllll@{}}
\toprule
parameters & $|H|$ & $p^{r,a}$ & $p^{r,y}$ & $b$ & $\sigma_{b}$ & $p_{sex}$   \begin{tabular}[c]{@{}l@{}}\end{tabular} \\ \midrule
default values      & 60      &    0.5   & 0.1  & 4    & 1     & 0.5                                               \\ \bottomrule
\end{tabular}
\end{table}

\begin{table*}
\footnotesize
\centering
\caption{Percentage of extinct and max limit runs. * indicates proportion of runs when the population reaches 500 agents.}
\label{tab:maxlimits}
\begin{tabular}{@{}lllllll@{}}
\toprule
          & \multicolumn{2}{c}{cubs}   & \multicolumn{2}{c}{yearlings} & \multicolumn{2}{c}{adults} \\ \midrule
scenarios & \% extinct & \% max limit* & \% extinct   & \% max limit*  & \% extinct & \% max limit* \\
+0.05     & 55\%       & 0\%           & 93\%         & 0\%            & 69\%       & 0\%           \\
+0.10     & 1\%        & 95\%          & 58\%         & 4\%            & 5\%        & 85\%          \\
+0.15     & 0\%        & 100\%         & 20\%         & 51\%           & 0\%        & 98\%          \\
+0.20     & 0\%        & 100\%         & 3\%          & 95\%           & 0\%        & 100\%         \\ \bottomrule
\end{tabular}
\end{table*}

\begin{figure*}
    \centering
    \caption{Average growth rate of population in different scenarios}
\begin{subfigure}[t]{0.49\textwidth}
\caption{Initial agent number scenarios}
\centering
\label{fig:initial}
\includegraphics[width=\textwidth]{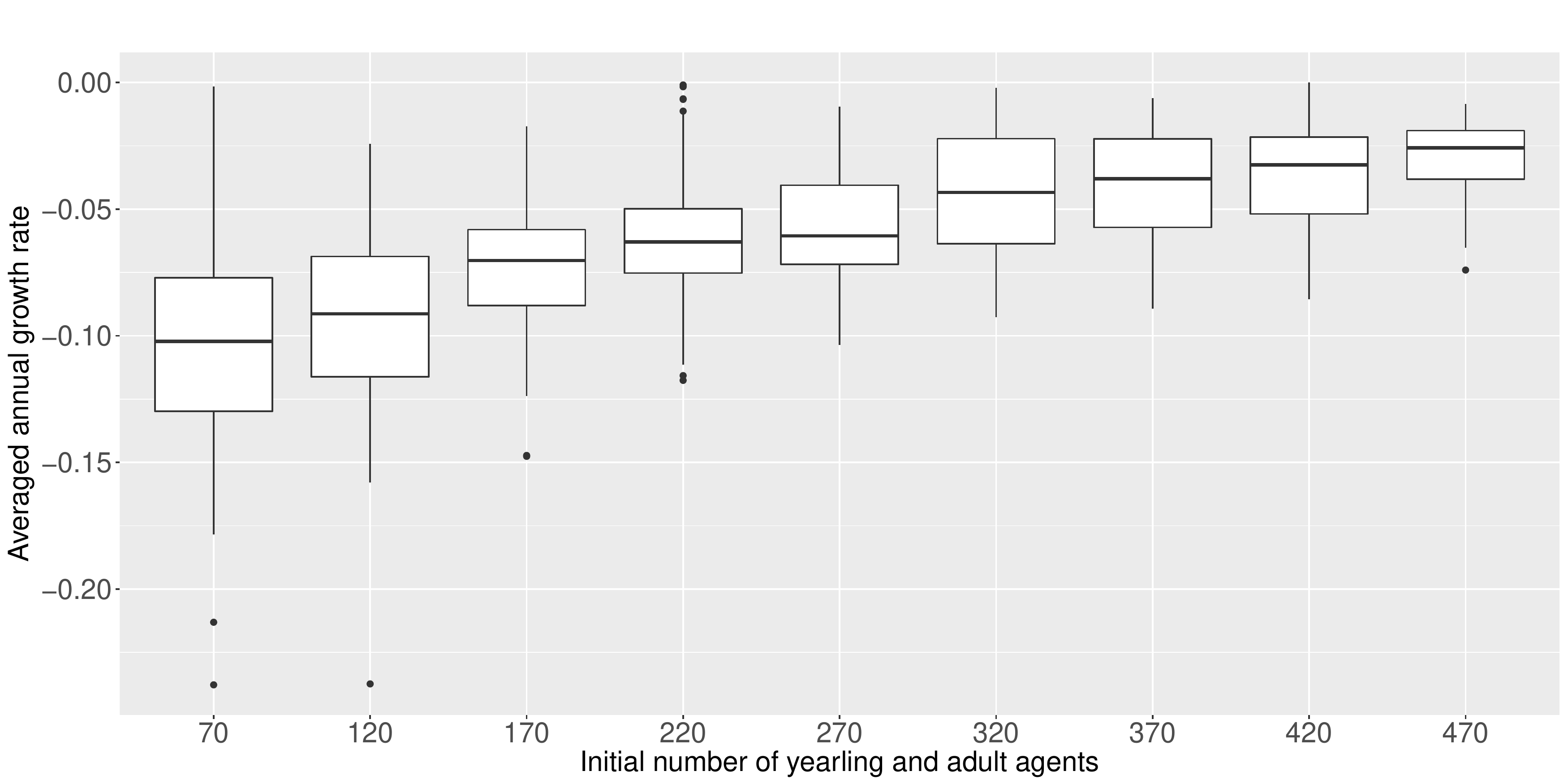}
\end{subfigure}
\begin{subfigure}[t]{0.49\textwidth}
\caption{Cub survival scenarios}
\centering
\label{fig:cubs}
\includegraphics[width=\textwidth]{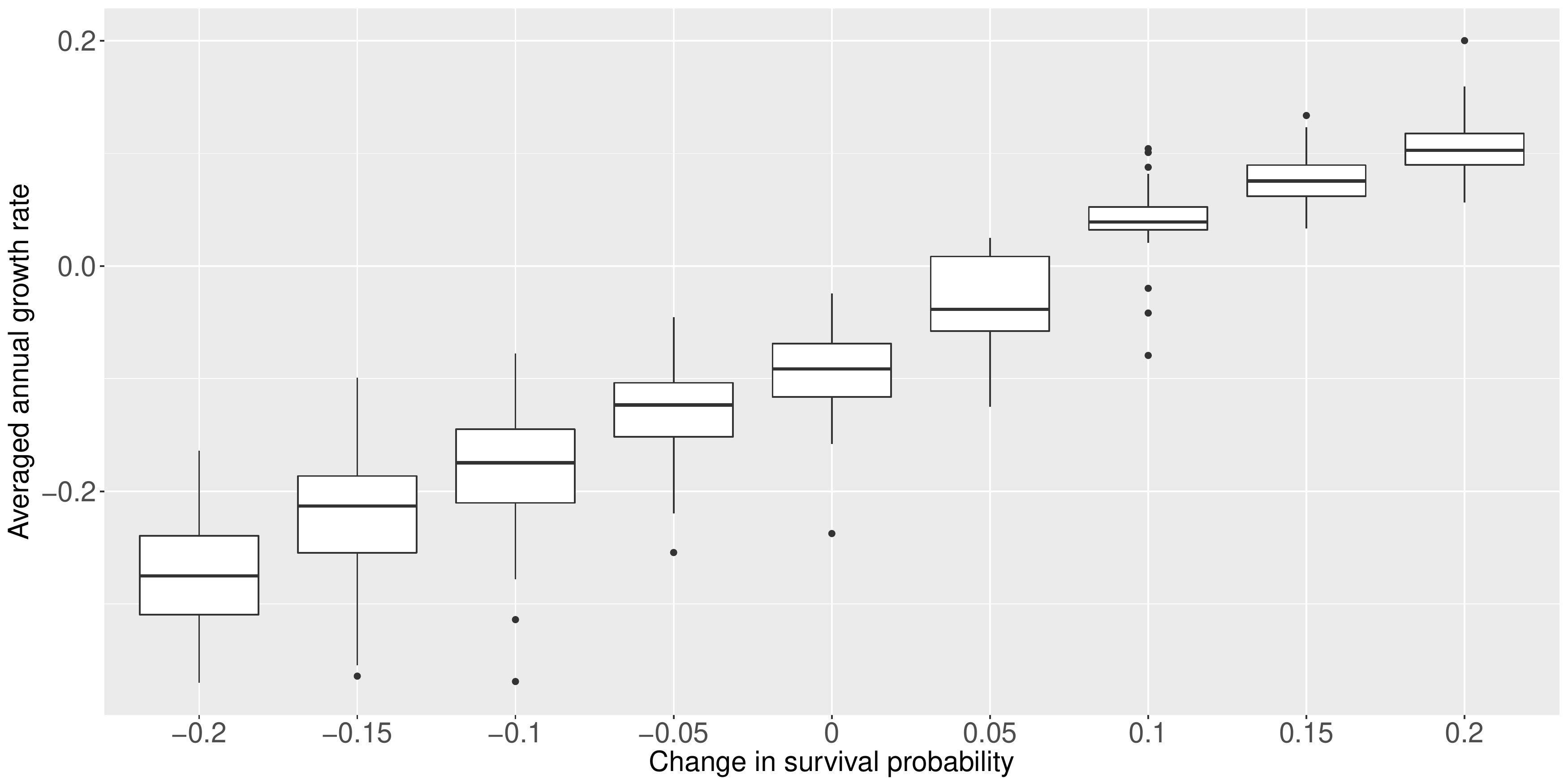}
\end{subfigure}
\end{figure*}

\subsection{Initialization}
At $t = 0$ cubs, yearlings, and adults are generated according to the following parameters: $n^{0}=120$ (yearling and adult total number), proportion of adult agents in initial population = 0.24, proportion of yearling agents = 0.15, proportion of cubs = 0.61. Sex is randomly assigned with 0.5 probabilities; age is assigned uniformly at random from 2 to 8 for adults (maximum observed age of Arctic foxes on Mednyi Island). Each agent is placed randomly on one of the home ranges.

\subsection{Empirical Simulations}
\textbf{Setup:} 
We implemented our model in Netlogo \cite{wilensky1999netlogo} and ran the model with default parameters while varying one of the parameters with others remained at default level (Table \ref{tab:overview}). The change of parameters in focus was performed in regular intervals ($\Delta$). We have completed 100 runs for each setting. Each run lasts 50 years or when the population went extinct. Because the computational costs rise substantially with the population size we limited the total number of agents number of agents excluding cubs ($n^{t}$) to 500 agents, which corresponds to the assumed ecological capacity of the island. The simulation stops upon reaching the maximum limit. In our analysis, only population-level variables were recorded, \emph{i.e.} the population size over time $n^{t}$, the averaged annual population growth ($\lambda$), and the proportion of runs with population reaching either extinction or the maximum limit. To allow the population reach its stationary age and sex composition \cite{gotelli1995primer} we excluded the first 3 years of each run from $\lambda$ calculation. We consider the population extinct when $n^{t}$ declines below the $10$ agent threshold.

\begin{table}
\footnotesize
\centering
\caption{Overview of parameters varying in simulations. $\Delta$ indicates the change in parameters.}
\label{tab:overview}
\begin{tabular}{@{}lll@{}}
\toprule
Parameter                        & Range                 & $\Delta$   \\ \midrule
$n^{0}$                        & 20-470                & 50   \\
Survival of cubs (both sex)      & $\pm 0.2$ of default value & 0.05 \\
Survival of yearlings (both sex) & $\pm 0.2$ of default value & 0.05 \\
Survival of adults (both sex)    & $\pm 0.2$ of default value & 0.05 \\ \bottomrule
\end{tabular}
\end{table}

\noindent\textbf{Critical mass for survival:} 
We focused on the impact of initial population of adults and yearling $n^{0}$ on the extinction of agents. 
In all scenarios, the population level significantly decreases, however the growth rate linearly rises with $n^{0}$ (Fig.\ref{fig:initial}). For larger $n^{0}$ scenarios ( $\geq 220$) the median for growth rates were similar ($\lambda = \{-0.024;-0.030\}$) at the initial stages of the simulation (first 10 years) and diverged afterwards. On average, the extinction  was equal 100\% for scenarios with $n^{0} \in \{20, 70 \}$, 99\% with $n^{0}=120$ (default) and gradually decreases as $n^{0}$ rises to 2\% for $n^{0} = 470$.

\noindent\textbf{Survival probabilities:} 
For all age class scenarios, $\lambda$ linearly increases with survival probabilities (e.g. see Fig.\ref{fig:cubs} for cub survival scenarios). The highest impact on population dynamics is affected first by cubs and then adults survival probabilities; a change of $+0.1$ and more to the default cub or adult survival probabilities leads to a rapid growth of population (Table \ref{tab:maxlimits}). For scenarios with $+0.05$ change to the default cub survival and $+0.1$ to the default yearling survival, trajectories of population dynamics significantly diverge between runs as average growth rate hovers around zero.

\section{Discussion}

We found that cub survival has the most pronounced impact on the population dynamic. It exceeds the effect of adult survival, though cub age class includes only one age cohort (0 year age) as compared to 11 for adults (2-12 year age). These results are consistent with the results of another individual-based model of Mednyi Arctic fox population \cite{ABM2018}, with Leslie matrix analysis of this population \cite{matrixMednyi2017} and with matrix modeling of other populations of ``fast'' (i.e, rapidly maturing) mammals \cite{van2013carnivora}.
The incorporation of AI techniques help reveal the threshold for the population size – around 200 adult and yearling agents -  below which the pace of decline accelerates. We attribute this effect to the growing influence of negative stochastic factors in small populations, \emph{i.e.} the result of demographic stochasticity \cite{morris2002quantitative}. We intend to investigate this effect in more details as a crucial dynamic factor for such small populations. 
The default parameters – aimed to replicate real characteristics of the population - result in a rapid extinction of agents, contradicting empirical evidence of the stable population dynamic for the past 20 years. Such an outcome may be attributed to failure to correctly identify default parameter values and/or simplistic structure of the model which does not incorporate some key factors and links. We note, that increasing the default parameter of the cub survival probability by $0.05$, \emph{i.e.} the magnitude of error in our estimation, improves the population dynamic with a growth rate increasing to $100$. The major shortcomings of our current model deal with the absence of (1) complete environmental factors, \emph{e.g.} spatial and temporal distribution of food, (2) spatial relationship between home ranges as a factor in dispersal submodel, and (3) the specific social structure of Mednyi foxes – females tend to stay at parental home ranges or move to adjacent ranges, so complex families with one male and several females emerge. We plan to incorporate these features during future development of the model extending further to the proposed final structure (Fig.\ref{fig:pgm}). 
In the future, we would like to study group formation and territory variations through techniques from cooperative and non-cooperative game theory: the behavior of Arctic foxes, particularly in forming families and packs, does not immediately correlate with food availability and reproductive abilities and seem to follow interesting, but non trivial patterns.

\begin{footnotesize}
\subsubsection*{Acknowledgments}
We are immensely grateful to Dr. Mikhail Goltsman, the chief of Medniy Arctic Fox project, who provided insights and expertise to this research. We thank our colleagues from Moscow State University, especially Elena Kruchenkova, who carried out field studies on Mednyi Island for over 20 years to collect empirical data used in this paper as well as Vladimir Burkanov and Nature and Biosphere Reserve of Commander Islands for their logistical support.
\end{footnotesize}

\bibliographystyle{named}
\bibliography{ref}

\end{document}